\documentclass[conference]{IEEEtran}
\IEEEoverridecommandlockouts
\usepackage{amsmath,amssymb,amsfonts}
\usepackage{algorithmic}
\usepackage{algorithm}
\usepackage{array}
\usepackage[caption=false,font=footnotesize,labelfont=rm,textfont=rm]{subfig}
\usepackage{textcomp}
\usepackage{stfloats}
\usepackage{url}
\usepackage{verbatim}
\usepackage{graphicx}
\usepackage{cite}
\usepackage{hyperref}
\usepackage{booktabs}
\usepackage{wrapfig}
\hypersetup{hypertex=true,
            colorlinks=true,
            linkcolor=blue,
            anchorcolor=blue,
            citecolor=blue}
\DeclareRobustCommand*{\IEEEauthorrefmark}[1]{%
    \raisebox{0pt}[0pt][0pt]{\textsuperscript{\footnotesize\ensuremath{#1}}}}

\def\BibTeX{{\rm B\kern-.05em{\sc i\kern-.025em b}\kern-.08em
    T\kern-.1667em\lower.7ex\hbox{E}\kern-.125emX}}
\begin{document}

\title{TRTAR: Transmissive RIS-assisted Through-the-wall Human Activity Recognition}

\author{
\IEEEauthorblockN{
Junshuo Liu\IEEEauthorrefmark{1}, Yunlong Huang\IEEEauthorrefmark{1}, Jianan Zhang\IEEEauthorrefmark{1}, Rujing Xiong\IEEEauthorrefmark{1}, 
Robert Caiming Qiu\IEEEauthorrefmark{1}}
\IEEEauthorblockA{\IEEEauthorrefmark{1}School of Electronic Information and Communications, Huazhong University of Science and Technology, Wuhan 430074, China}
Emails: \{junshuo\_liu, huangyunlong, zhangjn, rujing, caiming\}@hust.edu.cn
}

\maketitle

\begin{abstract}
Device-free human activity recognition plays a pivotal role in wireless sensing. However, current systems often fail to accommodate signal transmission through walls or necessitate dedicated noise removal algorithms. To overcome these limitations, we introduce TRTAR: a device-free passive human activity recognition system integrated with a transmissive reconfigurable intelligent surface (RIS). TRTAR eliminates the necessity for dedicated devices or noise removal algorithms, while specifically addressing signal propagation through walls. Unlike existing approaches, TRTAR solely employs a transmissive RIS at the transmitter or receiver without modifying the inherent hardware structure. Experimental results demonstrate that TRTAR attains an average accuracy of 98.13\% when signals traverse concrete walls.
\end{abstract}

\begin{IEEEkeywords}
Transmissive reconfigurable intelligent surface, human activity recognition, through-the-wall, machine learning.
\end{IEEEkeywords}

\section{Introduction}
Human activity recognition (HAR) stands as a prominent focus within the realms of computer vision and pattern recognition, serving as a pivotal technology in numerous applications, including human-computer interaction, health assessment, and elderly care \cite{hassan2018robust,zhou2020deep,zhao2022human}. Extensive research has been conducted on traditional vision-based and wearable device-based HAR methodologies \cite{xu2013exploring,kim2019vision,bhat2018online}. In comparison to these established sensing modalities, the wireless sensing-based approach exhibits robustness against variations in light conditions and environmental factors while safeguarding user privacy \cite{sun2022human}.

Recent advancements in wireless sensing-based HAR have extensively explored methodologies leveraging radio frequency identification (RFID), channel state information (CSI), and multi-antenna techniques to discern various activities. Noteworthy systems include TASA \cite{zhang2010tasa}, SmartWall \cite{oguntala2019smartwall}, WiSee \cite{pu2013whole}, CARM \cite{wang2015understanding}, RT-Fall \cite{wang2016rt}, WiAct \cite{yan2019wiact}, among others. SmartWall \cite{oguntala2019smartwall} utilized passive RFID tags for activity detection, while CARM \cite{wang2015understanding} employed CSI-speed and CSI-activity models for recognition. RT-Fall \cite{wang2016rt} utilized both amplitude and phase of CSI measurements for fall detection, and WiAct \cite{yan2019wiact} investigated correlations between body movement and CSI amplitude information for activity classification. Despite their advantages in passive detection and easy deployment, these systems have yet to address human activity recognition scenarios where wireless signals traverse through walls.

Indoor environments, particularly within home environments, often comprise multiple rooms accessing a single wireless access point (AP). In such settings, wireless signals necessitate passage through walls to reach their intended receivers. These walls can obstruct both direct and reflected propagation paths between the transmitter and receiver, significantly affecting signal transmission. Experimental evidence illustrates the substantial impact of walls on wireless signal propagation. For instance, 2.4 GHz Wi-Fi signals passing through an 18-inch-wide concrete wall experience an 18 dB attenuation \cite{adib2013see}. Our experiments further demonstrate that a 1.1-meter-thick concrete wall induces an 88.26 dB attenuation at 5.8 GHz. When all propagation paths are impeded by such walls, the environmental noise often eclipses subtle amplitude changes caused by user movements, consequently compromising the performance of activity recognition systems.

In this paper, we propose TRTAR, a passive human activity system utilizing a transmissive RIS with wireless signals, specifically tailored for scenarios involving signal transmission through walls. The primary technical challenge we confront in implementing this system is the acquisition of meaningful human activity CSI correlations from raw CSI measurements. This challenge is notably formidable due to the severe impact of concrete walls and indoor environmental factors on wireless signals. These factors lead to signal attenuation and introduce a complex background of environmental noise, making it arduous to discern human activity-induced changes in CSI waveforms when signals traverse walls. Prior endeavors have focused on applying denoising algorithms to address this challenge. For instance, Wu et al. \cite{wu2018tw} proposed an Or-PCA approach to establish correlations between human activity and CSI variations. Cao et al. \cite{cao2020research} utilized PCA and empirical mode decomposition (EMD) algorithms to mitigate wall interference and extract activity characteristics. In contrast to previous methodologies, our system tackles this challenge through hardware enhancements. The transmissive RIS is an almost passive electromagnetic (EM) material-based device deployable on various structures, including building facades and indoor walls, among others \cite{liu2021reconfigurable}. Importantly, the integration of the RIS does not alter the fundamental system architecture. By leveraging the transmissive RIS, our system significantly amplifies activity-related information without affecting the intrinsic system design, thereby substantially improving recognition accuracy.

The contributions of our work are summarized as follows:
\begin{itemize}
\item[$\bullet$] We propose a transmissive RIS-assisted human activity recognition system with wireless signals, which does not require any dedicated device and meets the scenarios of the signals through the wall. To the best of our knowledge, this is the first work to leverage the transmissive RIS for human activity recognition.
\item[$\bullet$] We implemented the transmissive RIS-assisted HAR system. Experiment results show that our system achieves an accuracy of 98.13\% when the signals pass through the concrete wall with the aid of the transmissive RIS.
\end{itemize}

The rest of the paper is structured as follows. Section II introduces a transmissive RIS prototype. Section III delineates the proposed HAR architecture, encompassing data collection methodologies, data analysis, and classification models. Experimental results and discussions are presented in Section IV. Finally, Section V encapsulates the drawn conclusions.

\section{PRELIMINARIES}
This section introduces the transmissive reconfigurable intelligent surface, signal attenuation model, and transmissive RIS-assisted signal enhancement.
\subsection{Transmissive Reconfigurable Intelligent Surfaces}
\begin{figure}[htbp]
\centerline{\includegraphics[width=0.9\columnwidth]{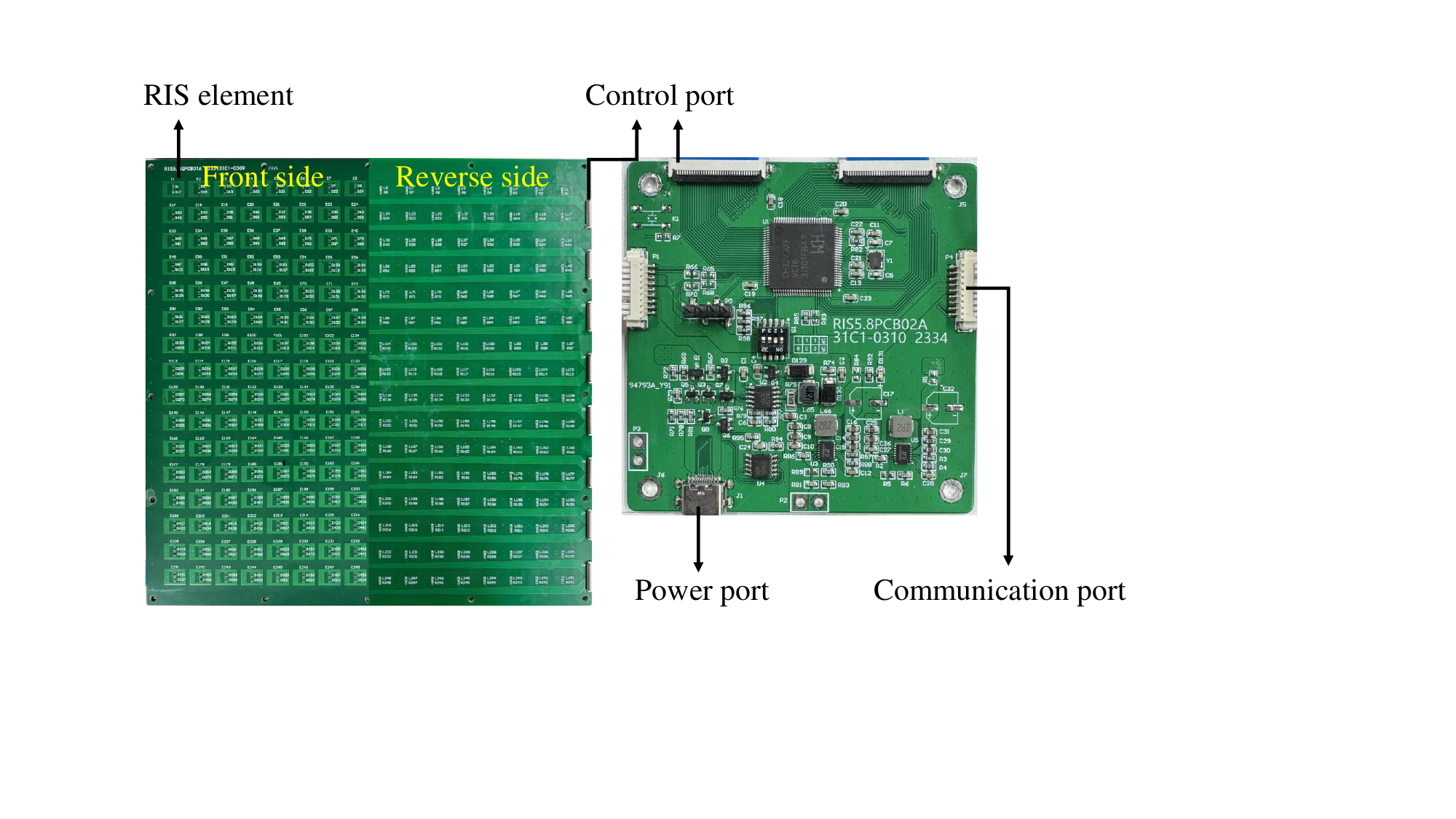}}
\caption{Photography of the 1-bit transmissive RIS prototype (left side), and the logic circuit controlling board (right side).}
\label{F1}
\end{figure}

A prototype of a 1-bit transmissive RIS operating at a frequency of 5.8 GHz is illustrated in Fig.~\ref{F1}. This RIS design, based on printed circuit board (PCB) technology, offers facile fabrication. Configured as a Uniform Planar Array (UPA), it comprises a grid consisting of dimensions $16 \times 16$ elements. The physical dimensions of this RIS prototype measure $31 \times 31 \, \text{cm}^2$. The primary objective entails achieving phase reconfigurability, necessitating individual control over each element within the array. To fulfill this requirement, 196 biasing lines are necessary, with each element associated with a single biasing line to apply DC voltages to two PIN diodes. To enable independent application of DC voltages across these 196 channels, four logic circuit controlling boards are affixed and interconnected with the biasing lines through multiple connectors.

\subsection{Wall Attenuation and Signal Enhancement}
We define path loss (PL) as the ratio between the effective transmitted power and the received power, accounting for system losses, amplifier gains, and antenna gains. Path loss concerning free space at a distance of 1 meter serves as a convenient reference for general link budget computations, formulated as shown in Equation \eqref{PL}:
\begin{align}
    P_R = & P_T + G_T + G_R - [\text{Path Loss w.r.t. 1 m FS}] \nonumber \\
          & + 20 \log_{10} \left ( \frac{\lambda}{4 \pi d}\right ),
\label{PL}
\end{align}
where $\lambda$ denotes the wavelength, while $G_T$ and $G_R$ represent the antenna gains of the transmitter and receiver, respectively, measured in decibels (dB). Furthermore, $P_T$ and $P_R$ stand for the transmitter and receiver powers, expressed in dBm \cite{rappaport2010wireless}.

Path loss can be described by the distance-dependent path loss model
\begin{equation}
\overline{PL}(d) [\text{dB}] = PL(d_0) [\text{dB}] + 10 n \log_{10} \left ( \frac{d}{d_0} \right ),
\end{equation}
where $\overline{PL}(d)$ represents the average path loss value in decibels (dB) at a separation distance $d$ between the transmitter and receiver (TR). $PL(d_0)$ denotes the path loss in dB at a reference distance $d_0 = 1 \text{m}$, while $n$ signifies the path loss exponent that delineates the rate of path loss escalation with increasing TR separation \cite{rappaport2010wireless}. For free space propagation, the value of $n$ is 2. However, real-world scenarios involving obstructions and multipath propagation alter the $n$ values in practical applications.

More sophisticated propagation models incorporate partition-dependent attenuation factors, presupposing a free path loss with $n = 2$, accompanied by supplementary path loss contingent upon the objects obstructing the space between the transmitter and the receiver. The partition-based path loss model is articulated as follows:
\begin{equation}
P_R = P_T + G_T + G_R + 20\log_{10} \left ( \frac{\lambda}{4 \pi d} \right ) - \sum_{i=1}^{N} \alpha_i,
\label{PLM}
\end{equation}
where $\alpha_i$ represents the attenuation value attributed to the $i$-th obstruction encountered along the path delineated by a line connecting the transmitter and the receiver point \cite{durgin1998measurements}.

\begin{figure}[tbp]
\centerline{\includegraphics[width=\columnwidth]{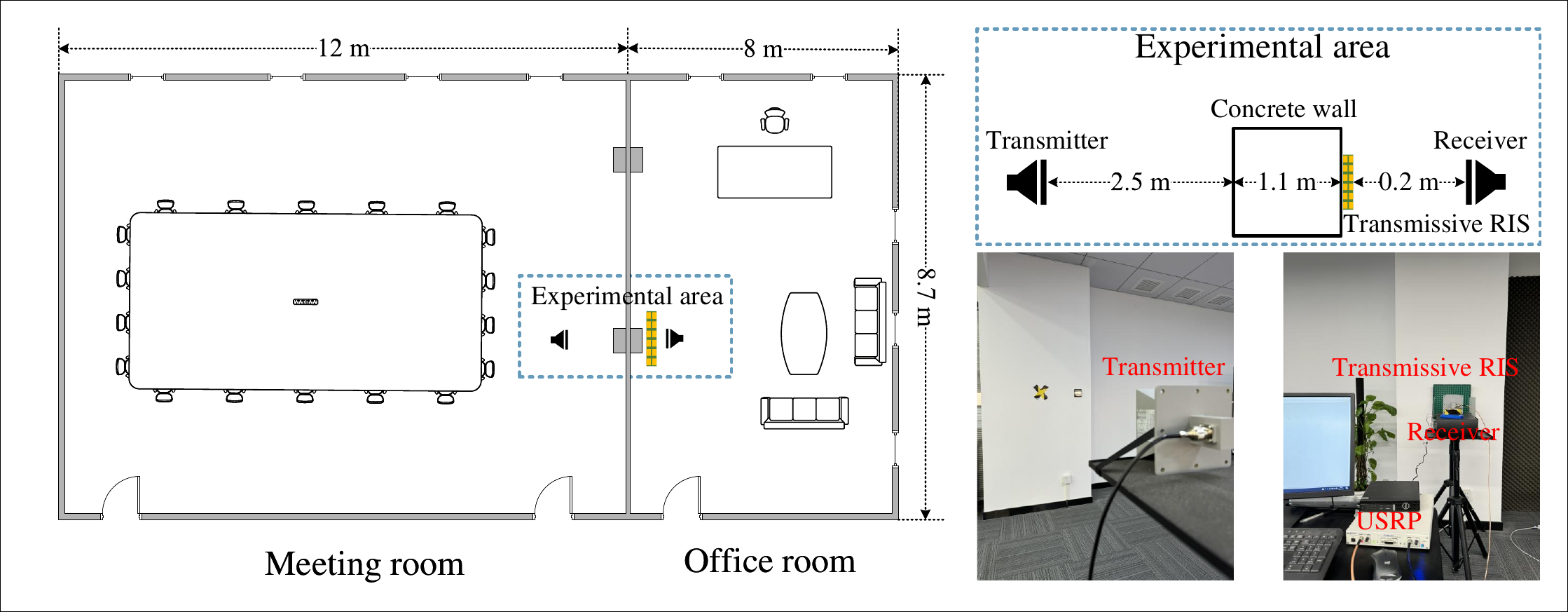}}
\caption{Floor plans of the experimental environment with concrete wall.}
\label{F3}
\end{figure}

To ascertain the effectiveness of the transmissive RIS, an experimental validation employing a known impulse response wired channel was established within our laboratory, illustrated in Fig.~\ref{F3}. The setup utilizes NI USRP-2954R hardware and Laboratory Virtual Instrument Engineering Workbench (LabVIEW) software. The propagation length within the validation scenario is fixed at 3.8 meters, wherein cable 1 spans 3 meters and cable 2 spans 10 meters. The transmitter horn antenna is linked to the USRP via cable 2, while the receiver horn antenna connects to the USRP via cable 1. Notably, the 13-meter cable exhibits an average attenuation of 16.51 dB at 5.8 GHz. Detailed parameters of the path loss measurements are outlined in Table~\ref{table1}.

\begin{table}[tbp]
\centering
\caption{Configurations in channel measurements.}
\label{table1}
\setlength{\tabcolsep}{4mm}
\begin{tabular}{ll}
\toprule
Parameters                 & Narrowband measurement
\\ 
\midrule
Frequency                  & 5.8 GHz                   
\\
Probing signal             & Continuous wave (CW)        
\\
Type of Tx/Rx antenna      & Directional horn antenna 
\\
Transmitted power          & 17 dBm                   
\\
Amplifier power            & 14 dB @ 5.8 GHz          
\\
Antenna gain               & 15.8 dBi @ 5.8 GHz          
\\
Height of Tx/Rx antenna    & 1 m                      
\\
Transmissive RIS dimension & $16 \times 16$
\\
\bottomrule
\end{tabular}
\end{table}

To assess signal attenuation owing to the concrete wall obstruction between the transmitter and receiver, an accurate characterization of the wall's electromagnetic properties within the experimental setting becomes imperative \cite{cuinas2001measuring}. Among these properties, the relative permittivity ($\varepsilon_r$) and conductivity ($\sigma$) exert the most substantial influence on signal attenuation \cite{pozar2011microwave}. Relative permittivity, often denoted as $\varepsilon_r(f)$, signifies the material's permittivity expressed as a ratio with the electric permittivity of a vacuum. It is fundamentally defined as
\begin{equation}
\varepsilon_r(f) = \frac{\varepsilon(f)}{\varepsilon_0},
\end{equation}
where $\varepsilon(f)$ is the complex frequency-dependent permittivity of the material, and $\varepsilon_0$ is the vacuum permittivity.

The propagation of radio waves encounters attenuation while traversing materials. While the expression for a general dielectric constant is intricate, a more straightforward evaluation is feasible within two limits: the dielectric limit as $\sigma \rightarrow 0$ and the good conductor limit as $\sigma \rightarrow \infty$ \cite{rudd2014building,series2015effects}:
\begin{equation}
\alpha = 1636 \frac{\sigma}{\sqrt{\varepsilon_r^{'}}},
\label{Attenuation}
\end{equation}
where $\varepsilon_r^{'}$ represents the real component of $\varepsilon_r$, while $\alpha$ signifies the specific material's attenuation value in dB/m. For concrete at 5.8 GHz, these parameters are $\varepsilon_r^{'}=3.58-5.50$ and $\sigma=0.11 \, \text{S/m}$ \cite{cuinas2000comparison,cuinas2001measuring,cuinas2002permittivity,cuinas2007modelling,ferreira2014review}. Consequently, considering Equation \eqref{PLM}, Equation \eqref{Attenuation}, and the details outlined in Table~\ref{table1}, the receiver power is computed to be -98.52 dBm.

\begin{figure}[htbp]
\centering
\subfloat[]{
\label{F4-a}
\includegraphics[width=0.5\columnwidth]{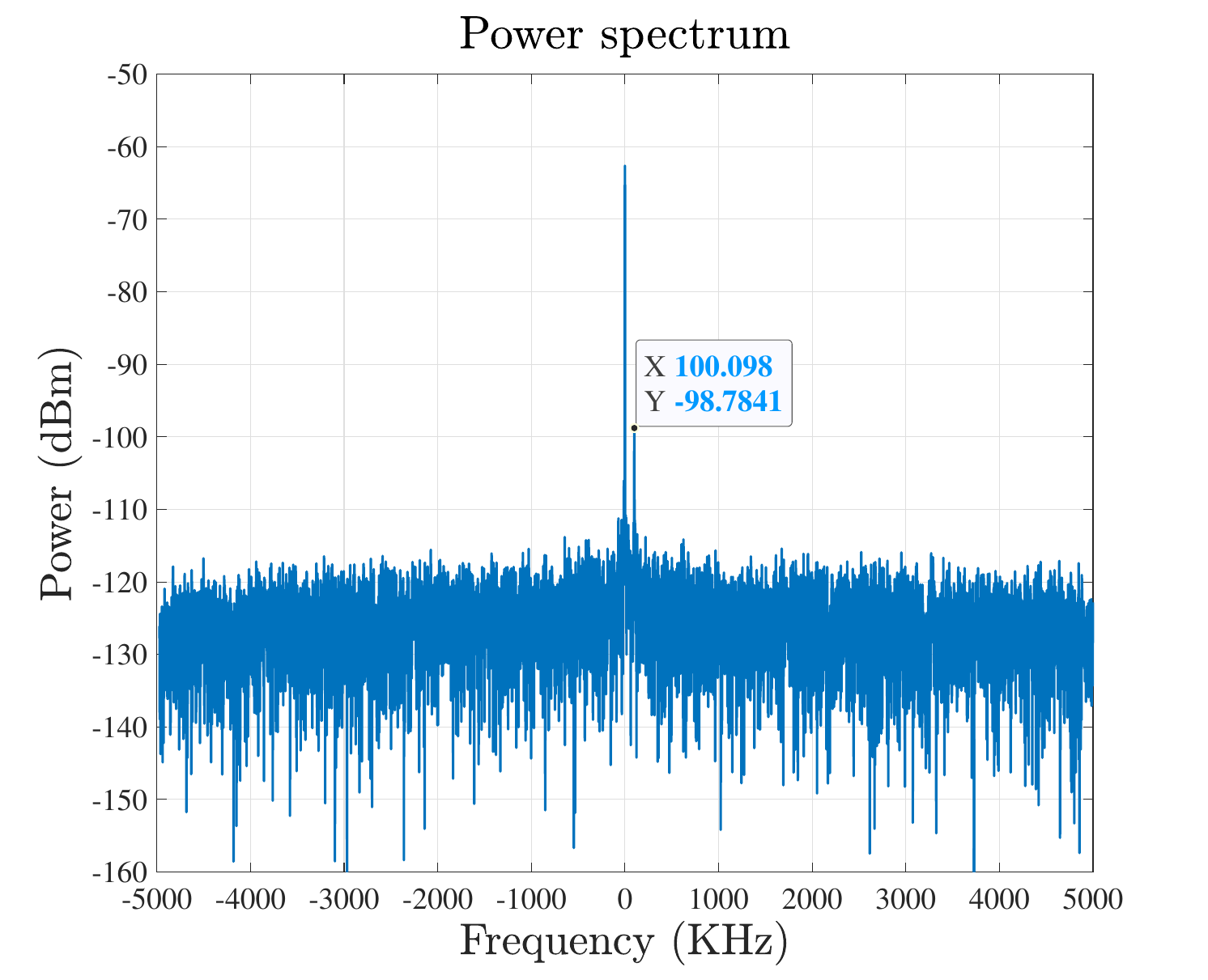}
}
\subfloat[]{
\label{F4-b}
\includegraphics[width=0.5\columnwidth]{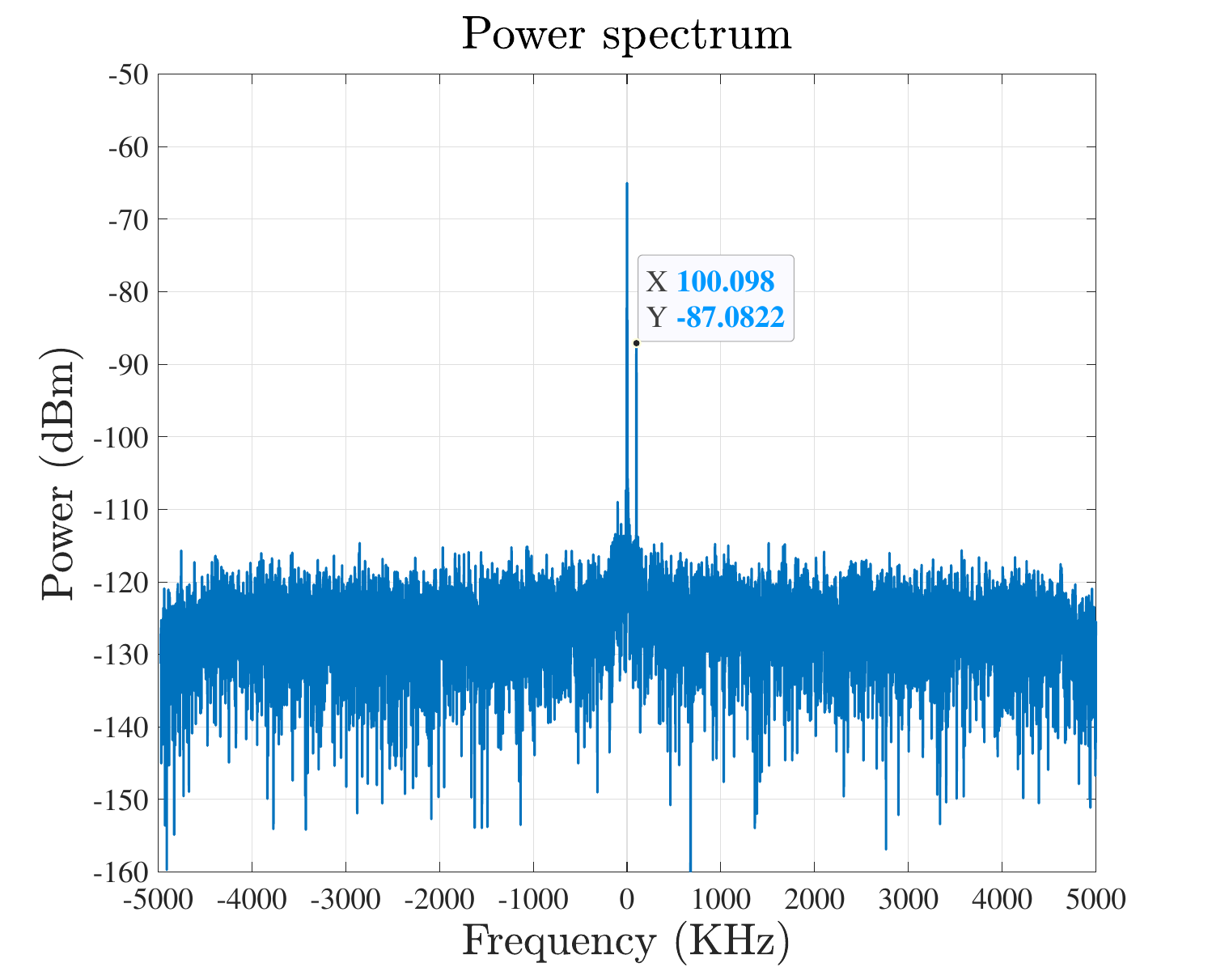}
}
\caption{The spectrum of the transmissive RIS test: (a) without RIS, (b) with RIS.}
\label{F4}
\end{figure}

In the validation, without the assistance of transmissive RIS, the recorded power at the horn antenna stands at -98.78 dBm. Upon integrating the transmissive RIS and scanning the formed beam towards the receiver horn antenna, a peak signal strength of -87.08 dBm is attained, as illustrated in Fig.~\ref{F4}~\subref{F4-b}. This observation demonstrates that the transmissive RIS yields an array gain of 11.7 dBi at 5.8 GHz. These experimental outcomes affirm the capacity of the transmissive RIS to enhance signal strength and effectively surmount obstacles along the TR path.

\section{System Framework}
\begin{figure}[htbp]
\centerline{\includegraphics[width=0.8\columnwidth]{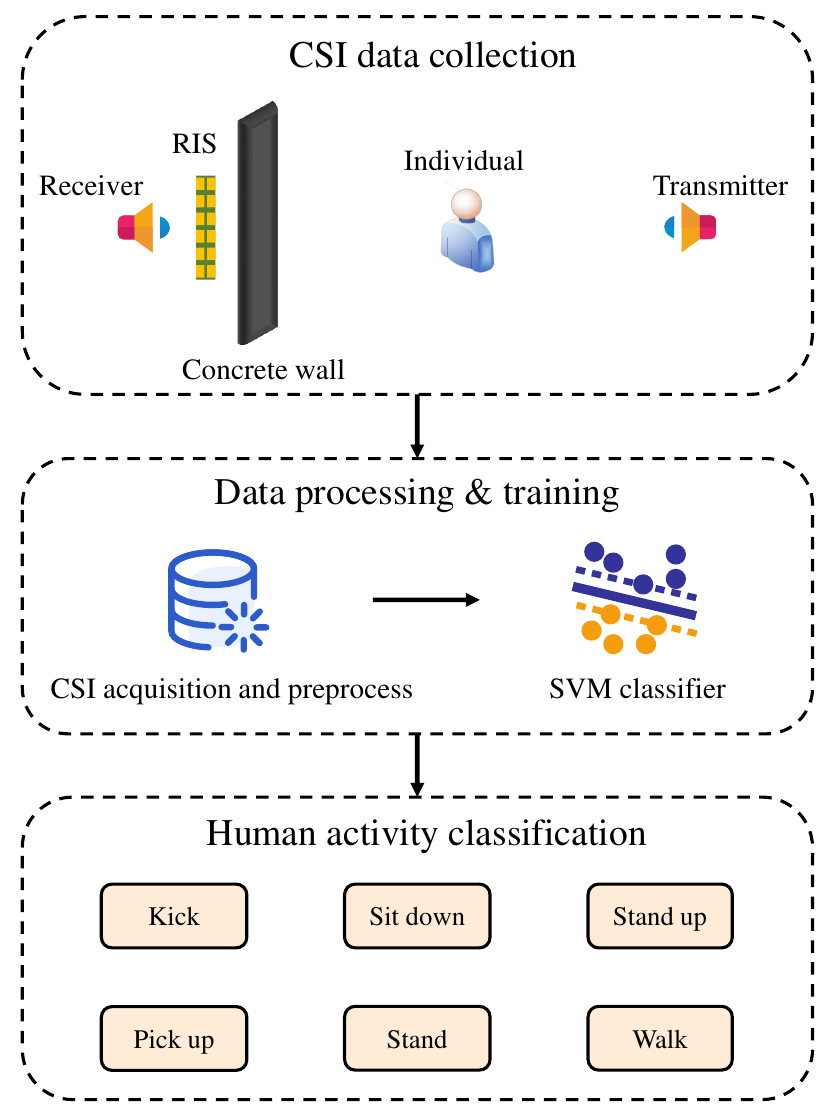}}
\caption{Overall architecture of the proposed HAR system.}
\label{F2}
\end{figure}

This section introduces the proposed system for human activity recognition, comprising three fundamental blocks: data collection, data preprocessing, and human activity recognition. Illustrated in Fig.~\ref{F2}, the utilization of a Universal Software Radio Peripheral (USRP) equipped with two horn antennas is deployed within the designated area to gather CSI data.

\subsection{Signal Analysis}
\begin{figure}[htbp]
\centering
\subfloat[Sit down]{
\label{F5-a}
\includegraphics[width=0.9\columnwidth]{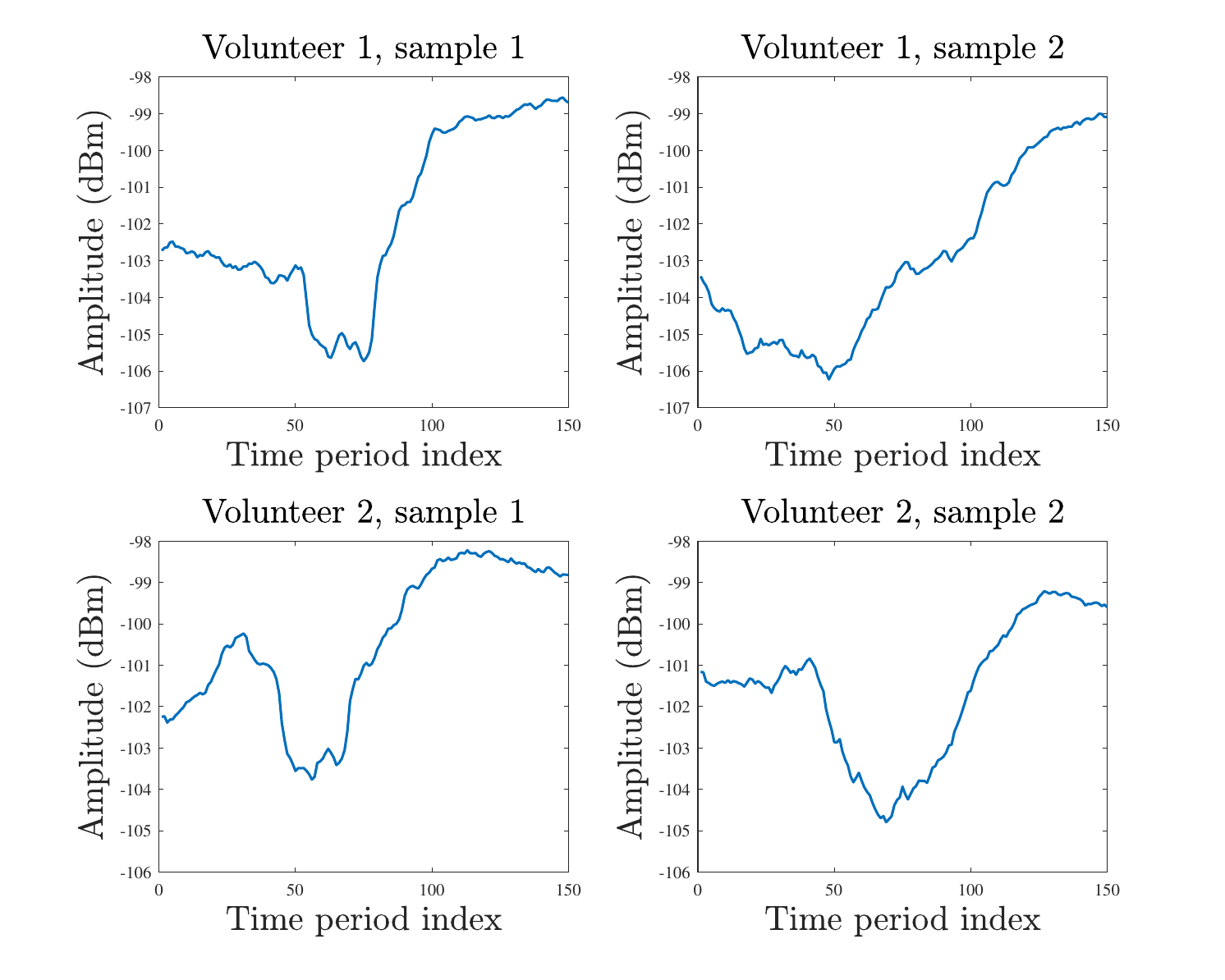}
}
\\
\subfloat[Walk]{
\label{F5-b}
\includegraphics[width=0.9\columnwidth]{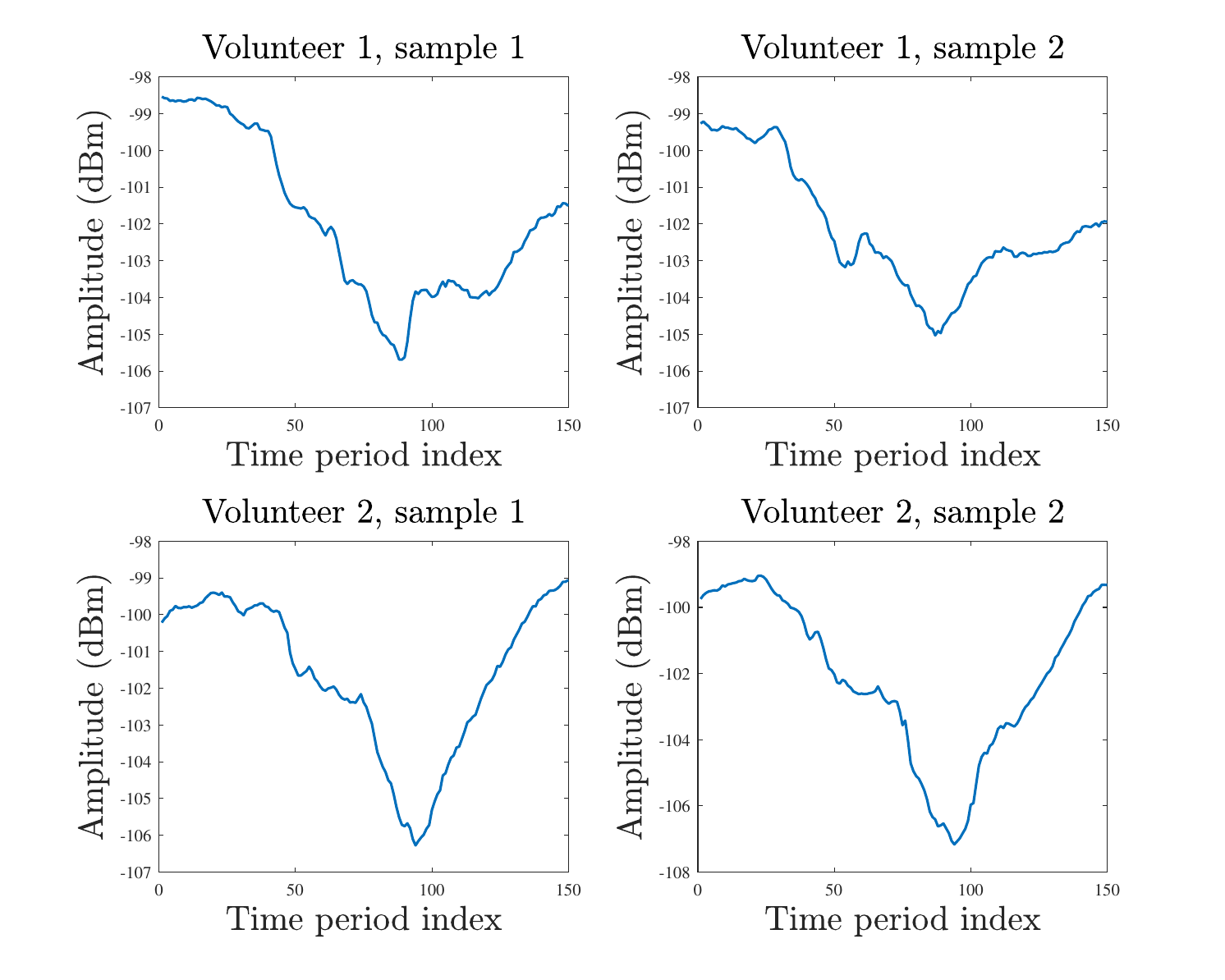}
}

\caption{The variation of the amplitude of CSI data when a person is performing specific activities.}
\label{F5}
\end{figure}

This subsection presents a visual representation of CSI samples corresponding to various activities. The plotted CSI samples are presented in the amplitude-versus-time format, wherein the vertical axis denotes the amplitude of the CSI samples and the horizontal axis represents the time duration. Fig.~\ref{F5} illustrates the CSI samples for different activities, revealing the comprehensive impact of distinct activities on the CSI amplitudes. The depicted figures showcase unique features within the CSI samples of various activities, substantiating the potential for activity recognition systems to distinguish between activities based on these received samples. Specifically, the real data visualization for the activities of sitting down and walking within the target area is delineated in Fig.~\ref{F5}~\subref{F5-a} and \subref{F5-b}. These plots offer a detailed observation of amplitude variations concerning individuals and activities, unveiling consistent trends in amplitude variations over time for similar activities. This consistency emphasizes that identical activities exhibit analogous trends within the time domain.

\subsection{Activity Recognition Methods}
In TRTAR, we employ the support vector machine (SVM), a machine learning classification algorithm, to discern human activities among the specified actions. SVM is regarded as a robust classifier method, displaying favorable performance compared to other classifier types. Its selection for this study is predicated on its efficacy in handling small datasets and its proficiency within high-dimensional spaces \cite{cortes1995support}.

\section{Experiments and Results}
\subsection{Experimental Setup}
The TRTAR system comprises a mini PC running the Windows 11 operating system and a NI USRP-2954R with two directional horn antennas serving as the transmitter and receiver components. Implementation of the TRTAR system involved configuring the mini PC with LabVIEW to collect CSI measurements, while MATLAB was employed for subsequent data processing. The USRP operated within the 5.8 GHz band with a 160 MHz channel bandwidth, and the CSI sampling rate was set at 20 Hz.

Evaluation of the TRTAR system's performance occurred within a specific environment depicted in Fig.~\ref{F3}. The environment consisted of a 1.1-meter concrete wall separating a meeting room, measuring 12 meters in length and 8.7 meters in width, from an adjoining office room measuring 8 meters in length and 8.7 meters in width. Deployment entailed placing the PC, NI USRP-2954R equipped with the receiving horn antenna, and transmissive RIS within the office room, while situating the transmitting horn antenna in the adjacent meeting room. The distance between the transmitter and receiver was 3.8 m.

\subsection{Performance Evaluation}
In this subsection, we evaluate the average cross-validation accuracy of the TRTAR system deployed between a meeting room and an office room. Four volunteers participated in collecting CSI data encompassing six distinct human activities: kicking, picking up, sitting down, standing, standing up, and walking. This spectrum of activities includes substantial movements such as walking and minor movements such as kicking. A total of 400 samples were gathered for each activity, resulting in a training dataset comprising 1680 samples and a testing dataset containing 720 samples. Subsequently, we extracted activity-specific features and trained the SVM model.

\begin{figure}[tbp]
\centerline{\includegraphics[width=0.9\columnwidth]{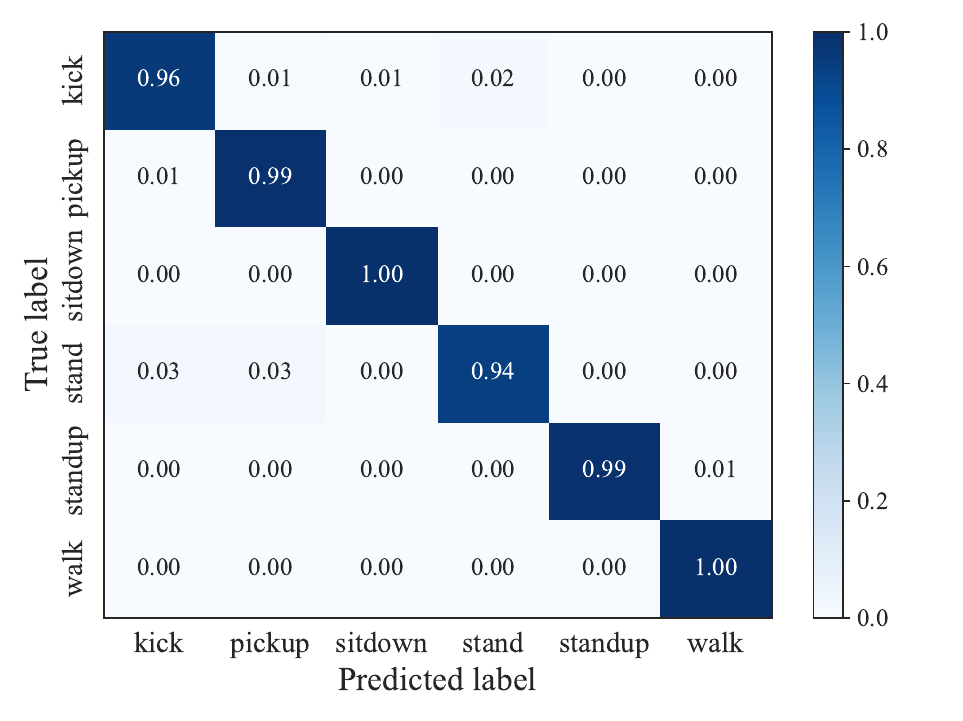}}
\caption{Experiment results of recognition accuracy with different human activities.}
\label{F6}
\end{figure}

Fig.~\ref{F6} depicts the 5-fold cross-validation human activity recognition accuracy within the meeting and office rooms using the TRTAR system. Across the spectrum of six human activities, TRTAR achieves an average cross-validation accuracy of 98.13\%. The individual accuracy rates for kicking, picking up, sitting down, standing, standing up, and walking are 96\%, 99.25\%, 100\%, 94.5\%, 99.25\%, and 100\%, respectively. Notably, our system attains 100\% accuracy in recognizing sitting down and walking, indicating an exclusive recognition of these activities without misclassifying them as other actions. Conversely, these activities also remain distinct from other recognized actions, reinforcing the system's discrimination capabilities. The experimental outcomes affirm the TRTAR's efficacy in achieving high accuracy across diverse human activities.

\section{Conclusion}
In this paper, we introduced TRTAR, a device-free passive human activity recognition system leveraging a transmissive RIS to accommodate signal transmission through walls. Unlike existing methodologies, TRTAR omits the use of additional noise removal algorithms. Instead, it employs the transmissive RIS to mitigate wall-induced interference, thereby enhancing hardware performance. The implementation of TRTAR involved the utilization of USRP devices and subsequent evaluation within an office environment. Our findings showcase that TRTAR achieved an average accuracy of 98.13\% in signal propagation through walls.

\section*{Acknowledgment}
This work was supported in part by the Nation Natural Science Foundation of China under Grant No.12141107, and in part by the Interdisciplinary Research Program of HUST, 2023JCYJ012.

\bibliographystyle{IEEEtran}
\bibliography{Reference}

\end{document}